\documentclass{KapProc} 

\normallatexbib

\begin{document}

\articletitle{Quintessence and Cosmic Microwave Background}

\author{Paul H. Frampton}
\affil{Department of Physics and Astronomy, University of North Carolina,\\
Chapel Hill, NC 27599-3255.}
\email{frampton@physics.unc.edu}

\begin{abstract}
Analytic formulas for the position of the first acoustic peak in
the CMB are derived and discussed. They are generalized to the case
of a time-dependent dark energy component and it is shown how the cosmic
parameters $\Omega_M$ and $\Lambda_{\Lambda}$, extracted from observations,
have an intrinsic uncertainty until one knows whether the dark energy
density is, or is not, time
dependent.
\end{abstract}

\bigskip

\section{CBR Temperature Anisotropy.}

Although the Cosmic Background Radiation (CBR) was first discovered
over thirty years ago \inxx{PW}, the detection of its temperature anisotropy
waited until 1992 when the Cosmic Background Explorer (COBE)
satellite provided its impressive experimental support\cite{smoot,ganga}
for the Big Bang model. In particular, the COBE results were consistent
with a scale-invariant spectrum of primordial scalar density
perturbations\cite{bar,sta,gupi,hawk} such as might be generated by quantum fluctuations
during an inflationary period.\cite{guth,lin,alb}

This discovery of temperature anisotropy in the CBR has
inspired many further experiments which will be sensitive to
smaller angle anisotropies than the COBE satellite was (about $1^o$).
NASA has approved the flight of a satellite mission, the Microwave
Anisotropy Probe (MAP) in the year 2000 and ESA has agreed to
a more accurate later experiment called the Planck Surveyor.
The expected precision of these measurements implies that
the angular dependence of the temperature anisotropy will
be known sufficiently well that the location of the first
acoustic (Doppler) peak, and possibly subsequent ones, will
be resolved. Actually, the BOOMERANG\cite{boomerang} data have already
provided a good measurement.

Although the hot big bang theory is supported by at least three
major triumphs: the expansion of the universe, the
cosmic background radiation and the nucleosynthesis calculations,
it leaves unanswered several questions. The most important
unanswered questions are the horizon and flatness issues.

When the CBR last scattered, the age of the universe was
about 100,000 years compared to its present age
of some 10 billion years. As we shall see, the horizon size
at the recombination time subtends now an angle
of about $(1/208)$  of $\pi$ radians. On the celestial sphere
there are therefore approximately 40,000 causally
disconnected regions. Nevertheless, these different
regions have a uniform CBR temperature to an accuracy
of better than one part in $10^5$. This is the
horizon problem.

The flatness problem may be understood from
the cosmological equation
\begin{equation}
\frac{k}{R^2} = (\Omega - 1) \frac{\dot{R}^2}{R^2}  
\label{cos}
\end{equation}
Evaluating Eq.(\ref{cos}) at an arbitrary time $t$ and dividing by the
same relation at the present time $t=t_0$ and using $ R \sim \sqrt{t}
\sim T^{-1}$ gives
\begin{equation}
(\Omega-1) = 4H_0^2t^2\frac{T^2}{T_0^2} (\Omega_0 -1)   
\label{omega}
\end{equation}
For high densities we write
\begin{equation}
\frac{\dot{R}^2}{R^2} = \frac{8 \pi G \rho}{3} = \frac{ 8 \pi Gga T^4}{6}
\end{equation}
where $a$ is the radiation constant and g is the effective number of degrees of freedom.
This leads to the relation between time and temperature, after
substituting the numerical values [$a=7.56\times10^{-9}erg m^{-3}K^{-4};
G/c^2 = 0.742\times 10^{-30}m/g; H_0=100h_0 km/s/Mpc = 3.25\times 10^{-18}h_0s^{-1}$]
\begin{equation}
t(seconds) = (2.42\times10^{-6})g^{-1/2}T^{-2}_{GeV}   
\label{time}
\end{equation}
Combining Eq.(\ref{omega}) with Eq.(\ref{time}) leads to
\begin{equation}
(\Omega - 1) = 3.64\times10^{-21}h_0^2g^{-1}T_{GeV}^{-2}(\Omega_0 - 1)
\end{equation}
Given the proximity of $\Omega_0$ to unity, we then deduce that
$\Omega$ at, for example, $T=1MeV$ ($t\sim$ 1second) must be
equal to one within one part in $10^{14}$! Otherwise the resultant
cosmology will be incompatible with the present situation
of our universe. This extraordinary fine-tuning is the flatness problem.

The goal\cite{stein1,stein2,stein3,kam1,kam2,kam3,kam4}
of the CBR experiments is to measure the temperature
autocorrelation function. The fractional temperature perturbation
as a function of the direction $\hat{{\bf n}}$ is expanded in
spherical harmonics
\begin{equation}
\frac{\Delta T(\hat{{\bf {n}}})}{T} = \sum_{lm} a_{(lm)}Y_{lm}(\hat{{\bf {n}}})
\end{equation}
and the statistical isotropy and homogeneity of the universe imply that
the coefficients have expectation values
\begin{equation}
<(a_{(lm)})^{*}a_{(l'm')}> = C_l\delta_{ll'}\delta_{mm'}
\end{equation}

The plot of $C_l$ versus $l$ is expected to reflect oscillations in
the baryon-photon fluid at the surface of last scatter. In particular,
the first Doppler peak should be at the postion $l_1 = \pi/\Delta\theta$
where $\Delta\theta$ is the angle now subtended by the horizon
at the time of the last scattering, namely the recombination time
corresponding to a red shift $z_t \sim 1,100$.

The horizon and flatness problems described above can both be solved
by the inflation scenario which has the further prediction
that $\Omega_0 = 1$ if the cosmological constant vanishes or
more generally that $\Omega_0 + \Omega_{\Lambda} = 1$ if
the cosmological constant does not vanish.

The question we address here is restricted to the question
of how much the value of $l_1$ alone - likely to be
accurately determined in the next few years -  will
tell us about the values of the cosmic parameters $\Omega_0$
and $\Omega_{\Lambda}$?

In Section 2, the case $\Lambda=0$ is discussed. In Section 3,
there is the more general case; in Section 4
there is discussion of the Figures derived;
finally, in Section 5, there is an amplification of the cosmological
constant and its generalization to quintessence..

\bigskip

\section{The Special Case $\Lambda=0$, $0 < \Omega_0 < 1$}

When the cosmological constant vanishes, the Einstein-Friedmann
cosmological equations can be solved analytically
(not the case, in general, when $\Lambda\neq0$). So we shall begin
by doing this special case explicitly. It gives rise
to the well-known result that the position of the first
Doppler peak (partial wave $l_1$) expected in the partial-wave
analysis depending on the present matter-energy density
$\Omega_0$ (for $\Lambda = 0$) according to $l_1 \sim 1/\sqrt{\Omega_0}$
\cite{stein3,kam4}.  We shall show in the next
section how in the general case with $\Lambda\neq0$ there is a
rather serious "comic confusion" in disentangling the value
of $\Omega_0$ from the position $l_1$ of the first Doppler peak.
Let us use the metric:

\begin{equation}
ds^2 = dt^2 - R^2[d\Psi^2 + sinh^2\Psi d\theta^2 + sinh^2\Psi sin^2\theta d\phi^2]
\label{metric}
\end{equation}
For a geodesic $ds^2=0$ and, in particular,

\begin{equation}
\frac{d\Psi}{dR} = \frac{1}{R}
\end{equation}

\noindent Einstein's equation reads

\begin{equation}
\left( \frac{\dot{R}}{R} \right)^2 = \frac{8\pi}{3}G\rho + \frac{1}{R^2}
\label{einstein}
\end{equation}
where we take curvature $k=-1$. Let us define:

\begin{equation}
\Omega_0 = \frac{8 \pi G \rho_0}{3H_0^2}; \rho=\rho_0\left(\frac{R_0}{R}\right)^3; a=\Omega_0H_0^2R_0^3
\end{equation}
Then from Eq.(\ref{einstein}) we find that

\begin{equation}
\dot{R}^2R^2 = R^2 + aR
\label{RR}
\end{equation}
and so it follows that

\begin{equation}
\frac{d\Psi}{dR} = \frac{d\Psi}{dt}\left(\frac{dR}{dt}\right)^{-1} =
\frac{1}{\dot{R}R} = \frac{1}{\sqrt{R^2 + aR}}
\end{equation}
Since $\Psi_0 = 0$, the value at time $t$ can be computed from the integral

\begin{equation}
\Psi_t = \int_{R_t}^{R_0} \frac{dR}{\sqrt{(R + a/2)^2 - (a/2)^2}}
\label{integral}
\end{equation}
This can be performed easily with the substitution
$R = \frac{1}{2}a(coshV - 1)$ to give the result:
\begin{equation}
\Psi_t = cosh^{-1} \left( \frac{2R_0}{a} - 1\right) -
cosh^{-1} \left( \frac{2R_t}{a} - 1\right)
\label{Psi}
\end{equation}
From Eq.(\ref{einstein}) evaluated at $t=t_0$ we see that
\begin{equation}
\frac{1}{a} = \frac{1 - \Omega_0}{R_0 \Omega_0}
\end{equation}
and so, using $sinh(cosh^{-1}x)=\sqrt{x^2-1}$ in Eq.(\ref{Psi}) gives now
\begin{equation}
sinh\Psi_t = \sqrt{\left(\frac{2(1-\Omega_0)}{\Omega_0} + 1\right)^2 - 1} -
\sqrt{\left(\frac{2(1-\Omega_0)R_t}{\Omega_0R_0} + 1\right)^2 - 1}
\label{sinh}
\end{equation}
The position of the first Doppler peak depends on the angle subtended
by the horizon size at the time $t$ equal to the recombination time.
This corresponds to the distance $(H_t)^{-1}$. According to the metric
of Eq.(\ref{metric}) the angle subtended is
\begin{equation}
\Delta \theta = \frac{1}{H_t R_t sinh \Psi_t}     
\label{theta}
\end{equation}
and the position of the first Doppler peak corresponds to the partial
wave $l_1$ given by
\begin{equation}
l_1 = \frac{\pi}{\Delta \theta} = \pi H_t R_t sinh \Psi_t    
\label{l}
\end{equation}
Now the red-shift at recombination is about $z_t=1,100 \simeq (R_0/R_t) \gg 1$
so we may approximate in Eq.(\ref{sinh}) to find
\begin{equation}
sinh\Psi_t \simeq \frac{2 \sqrt{1 - \Omega_0}}{\Omega_0}
\end{equation}
Using $H_t^2 = 8\pi G \rho/3 + 1/R^2 \simeq \Omega_0h_0^2(R_0/R)^3$
gives
\begin{equation}
l_1(\Lambda = 0) = \frac{2 \pi}{\sqrt{\Omega_0}} z_t^{1/2}
\end{equation}
In particular, if $\Omega_0 = 1$ and $\Lambda = 0$, one has $l_1 \simeq 208.4$.
If $l_1$ does have this value empirically it will favor this
simplest choice, although as we shall see in the following subsection
even here the conclusion has ambiguities.

In Fig. 1 of \cite{FNR} is plotted $l_1$ versus $\Omega_0$ for the particular case of
$\Omega_{\Lambda} = 0$.

\bigskip
\bigskip

\section{The General Case: $0 \leq \Omega_0 < 2; 0 \leq \Omega_{\Lambda} < 1$}

For the general case of $0 \leq \Omega_{\Lambda} < 2; 0 < \Omega_0 < 1$
we use the more general Einstein cosmological equation:
\begin{equation}
\dot{R}^2R^2 = -k R^2 + aR + \Lambda R^4 /3   
\label{geneinst}
\end{equation}
where $a = \Omega_0 H_0^2 R_0^3$. We define
\begin{equation}
\Omega_0 = \frac{8 \pi G \rho_0}{3 H_0^2}; \Omega_{\Lambda} =
\frac{\Lambda}{3 H_0^2}; \Omega_C = \frac{-k}{H_0^2 R_0^2}
\end{equation}
Substituting $R = R_0 r$ and $w = 1/r$ now gives rise to the integral\cite{int} for $\Psi_t$
\begin{equation}
\Psi_t = \sqrt{\Omega_C} \int_1^{\infty} \frac{dw}{\sqrt{\Omega_{\Lambda} + \Omega_C w^2 + \Omega_0 w^4}}
\end{equation}
in which $\Omega_{\Lambda} + \Omega_{C} + \Omega _0 = 1$.

Consider first the case of an open universe $\Omega_C > 0$. Then
\begin{equation}
l_1 = \pi H_t R_t sinh\Psi_t
\end{equation}

We know that
\begin{eqnarray}
H_t^2 & = & \left( \frac{\dot{R}_t}{R_t}
\right)^2 = \frac{8 \pi G \rho}{3} + \frac{\Lambda}{3} + \frac{1}{R_t^2} \nonumber \\
& = & H_0^2 \left[ \Omega \left( \frac{R_0}{R_t}\right)^3 +
H_0^2 \Omega_{\Lambda} + \left( \frac{R_0}{R_t}\right)^2 \Omega_C \right]
\end{eqnarray}
Since $R_0 \gg R_t$ we may approximate:
\begin{equation}
H_t \simeq \left( \frac{R_0}{R_t} \right)^{3/2} H_0 \sqrt{\Omega_0}
\end{equation}
and hence
\begin{equation}
H_tR_t = \left( \frac{R_0}{R_t} \right)^{1/2} \sqrt{\frac{\Omega_0}{\Omega_C}}
\end{equation}
It follows that for this case $\Omega_C > 0$ that
\begin{equation}
l_1 = \pi \sqrt{\frac{\Omega_0}{\Omega_C}}
\left(\frac{R_0}{R_t}\right)^{1/2} sinh \left( \sqrt{\Omega_C} \int_1^{\infty}
\frac{dw}{\sqrt{\Omega_{\Lambda} + \Omega_C w^2 + \Omega_0 w^3}} \right)
\label{l1c+}
\end{equation}
For the case $\Omega_C < 0 (k = +1)$ we simply replace $sinh$ by $sin$ in Eq. (\ref{l1c+}).
Finally, for the special case $\Omega_C = 0$, the generalized flat case favored by
inflationary cosmologies, Eq.(\ref{l1c+}) simplifies to:
\begin{equation}
l_1 = \pi \sqrt{\Omega_0}
\left(\frac{R_0}{R_t}\right)^{1/2}  \int_1^{\infty}
\frac{dw}{\sqrt{\Omega_{\Lambda} + \Omega_0 w^3}}
\end{equation}
In Fig. 2 of \cite{FNR} is plotted the value of $l_1$ versus $\Omega_0$ for
the case $\Omega_C = 0$ (flat spacetime). The contrast with
Fig 1 of \cite{FNR} is clear: whereas $l_1$ increases with decreasing $\Omega_0$
when $\Omega_{\Lambda}=0$ (Fig. 1 of \cite{FNR}) the opposite behaviour occurs when
we constrain $\Omega_{\Lambda} = 1 - \Omega_0$ (fig.2 of \cite{FNR}).

With $\Omega_0$ and $\Omega_{\Lambda}$ unrestricted there are more
general results. In Fig. 3 of \cite{FNR} are displayed iso-l lines on a
$\Omega_0 - \Omega_{\Lambda}$ plot.
The iso-l lines are (from right to left) for the values
$l_1 = 150, 160, 170, 180, 190, 200, 210, 220,\\
 230, 240, 250, 260, 270$ respectively.
One can see that from the {\it position}($l_1$)  only of the first Doppler peak
there remains a serious ambiguity of interpretation without further information.

In Fig, 4 of \cite{FNR}, there is a three dimensional rendition of the
value of $l_1$ versus the two variables $\Omega_0$ and $\Omega_{\Lambda}$.

\bigskip
\bigskip

\section{Discussion of Cosmic Parameter Ambiguities.}

Let us now turn to an interpretation of our Figures, from
the point of view of determining the cosmic parameters.

In the case where $\Lambda = \Omega_{\Lambda}=0$, Fig.1
of \cite{FNR} is
sufficient. In this case, there is the well-known dependence\cite{stein3,kam4}
$l_1 = (208.4)/\sqrt{\Omega_0}$ illustrated in Fig.1.
It would be straightforward to determine $\Omega_0$ with
an accuracy of a few percent from the upcoming measurements.

Of course there is a strong theoretical prejudice towards
$\Lambda=0$. But no underlying symmetry principle is yet
known. If $\Omega_{\Lambda} \neq 0$, one knows that it
is not bigger than order one; this is very many orders
of magnitude smaller than expected\cite{WN} from the vacuum energy
arising in spontaneous breaking of symmetries such as the electroweak
group $SU(2) \times U(1)$.

Nevertheless, recent observations of high redshift Type 1a supernovae
have led to the suggestion of an {\it increasing} Hubble parameter
\cite{perl1,perl2}. An interpretation of this is that
the cosmological constant is non-zero, possibly
$\Omega_{\Lambda} \simeq 0.7$ but is still consistent
with $\Omega_0 = 1 - \Omega_{\Lambda}$.
These results are certainly enough to motivate
a full consideration of non-zero values of $\Omega_{\Lambda}$.

Thus we come to Fig. 2 of \cite{FNR} which depicts the $\Omega_0$ dependence
of $l_1$ when $\Omega_0 + \Omega_{\Lambda} = 1$ is held fixed
as in a generalized flat cosmology that could arise from inflation.
We notice that here $l_1$ {\it decreases} as $\Omega_0$
decreases from $\Omega_0 = 1$, the opposite behaviour
to Fig. 1 of \cite{FNR}. Thus even the shift of $l_1$ from $l_1 =208.4$
depends on the size of $\Lambda$.

It is therefore of interest to find what are the contours
of constant $l_1$ in the $\Omega_0 - \Omega_{\Lambda}$
plane. These iso-l lines are shown in Fig. 3 of \cite{FNR}for
$l_1 = 150,....,270$ in increments $\Delta l_1 = 10$.
If we focus on the $l_1 = 210$ contour (the seventh contour from
the left in Fig. 3 of \cite{FNR}) as an example,
we see that while this passes close to the
$\Omega_0 = 1, \Lambda = 0$ point it also tracks out
a line naturally between those shown in Figs. 1 and 2 (actually
somewhat closer to the latter than the former).

Fig. 4  of \cite{FNR} gives a three-dimensional rendition which
includes the Figures 1-3 of \cite{FNR} as special cases and provides a visualisation
of the full functional dependence of $l_1(\Omega_0, \Omega_{\Lambda})$.

Our main conclusion is that the position $l_1$ of the first
Doppler peak will define the correct contour in our
iso-l plot, Fig. 3 of \cite{FNR}. More information will be necessary
to determine $\Omega_0$ and the validity of inflation.

\bigskip
\bigskip

\section{The Cosmological Constant Reconsidered.}

\bigskip
\bigskip

\noindent Our knowledge of the universe has changed dramatically even in the
last few years. Not long ago the best guess, inspired partially by inflation,
 for the makeup of the present cosmological energy density
was $\Omega_m = 1$ and $\Omega_{\Lambda} = 0$. However, the recent
experimental data on the cosmic background radiation and the
high - $Z$ ($Z$ = red shift) supernovae strongly suggest that both
guesses were wrong. Firstly $\Omega_m \simeq 0.3 \pm 0.1$. Second,
and more surprisingly, $\Omega_{\Lambda} \simeq 0.7 \pm 0.2$.
The value of $\Omega_{\Lambda}$ is especially unexpected for two reasons:
it is non-zero and it is $\geq 120$ orders of magnitude below its ``natural''
value.

\bigskip

The fact that the present values of $\Omega_m$ and $\Omega_{\Lambda}$
are of comparable order of magnitude is a ``cosmic coincidence''
if $\Lambda$ in the Einstein equation

\[ R_{\mu\nu} - \frac{1}{2} g_{\mu\nu} R = 8 \pi G_N T_{\mu\nu} + \Lambda g_{\mu\nu}
\]

\noindent is constant. Extrapolate the present values
of $\Omega_m$ and $\Omega_{\Lambda}$ back, say, to redshift
$Z = 100$. Suppose for simplicity that the universe is flat
$\Omega_C = 0$ and that the present cosmic parameter values are
$\Omega_m = 0.300...$ exactly and $\Omega_{\Lambda} = 0.700...$ exactly.
Then since $\rho_m \propto R(t)^{-3}$ (we can safely neglect radiation),
we find that
$\Omega_m \simeq 0.9999..$ and $\Omega_{\Lambda} \simeq 0.0000..$ at
$Z = 100$. At earlier times the ratio $\Omega_{\Lambda} /\Omega_m$
becomes infinitesimal.
There is nothing to exclude these values but it does introduce
a second ``flatness'' problem because, although we can argue for $\Omega_m + \Omega_{\Lambda}
= 1$ from inflation, the comparability of the present values of
$\Omega_m$ and $\Omega_{\Lambda}$ cries out for explanation.

\bigskip

In the present Section 5 we shall 
consider a specific model of quintessence. In its context
we shall investigate the position of the first Doppler peak in the Cosmic
Microwave Background (CMB) analysis using results published by two of us with Rohm
earlier\cite{FNR}. Other works on the study of CMB include\cite{K,B,BTW,LSW}.
We shall explain some subtleties of the derivation given
in \cite{FNR} that have been raised since its publication mainly because the
formula works far better than its expected order-of-magnitude accuracy.
Data on the CMB have been provided recently in
\cite{L,DK,M+,PSW,E,TZ,L+,l+} and especially in \cite{boomerang}.

The combination of the information about the first Doppler peak and the complementary
analysis of the deceleration parameter derived from observations of the high-red-shift
supernovae\cite{Perlmutter,Kirshner} leads to fairly precise values for the cosmic
parameters $\Omega_m$ and $\Omega_{\Lambda}$. We shall therefore also investigate the
effect of quintessence on the values of these parameters.

In \cite{FNR}, by studying the geodesics in the post-recombination period a formula was arrived
at for the position of the first Doppler peak, $l_1$.
For example, in the case of a flat universe with $\Omega_C = 0$
and $\Omega_M + \Omega_{\Lambda} = 1$ and for a conventional cosmological constant:

\bigskip

\begin{equation}
l_1 = \pi \left( \frac{R_t}{R_0} \right)
\left[\Omega_M  \left( \frac{R_0}{R_t} \right)^3 + \Omega_{\Lambda} \right]^{1/2}
\int_1^{\frac{R_0}{R_t}} \frac{dw}{\sqrt{\Omega_M w^3 + \Omega_{\Lambda}}}
\label{l1flat}
\end{equation}

\bigskip

\noindent If $\Omega_{C} < 0$ the formula becomes

\bigskip

\begin{eqnarray}
l_1 & = & \frac {\pi}{\sqrt{-\Omega_C}} \left( \frac{R_t}{R_0} \right)
\left[\Omega_M  \left( \frac{R_0}{R_t} \right)^3 + \Omega_{\Lambda} + \Omega_C
\left( \frac{R_0}{R_t} \right)^2 \right]^{1/2} \times \nonumber \\
& & \times ~~~ {\rm sin} 
\left( \sqrt{-\Omega_C} \int_1^{\frac{R_0}{R_t}} \frac{dw}{\sqrt{\Omega_M w^3 + \Omega_{\Lambda}}}
\right)
\label{l1open}
\end{eqnarray}

\bigskip

\noindent For the third possibility of a closed universe with $\Omega_C > 0$ the formula
is:

\bigskip

\begin{eqnarray}
l_1 & = & \frac {\pi}{\sqrt{\Omega_C}} \left( \frac{R_t}{R_0} \right)
\left[\Omega_M  \left( \frac{R_0}{R_t} \right)^3 + \Omega_{\Lambda} +5 \Omega_C
\left( \frac{R_0}{R_t} \right)^2 \right]^{1/2} \times \nonumber \\
& & \times ~~~
{\rm sinh} \left( \sqrt{\Omega_C} \int_1^{\frac{R_0}{R_t}} \frac{dw}{\sqrt{\Omega_M w^3 + \Omega_{\Lambda}}}
\right)
\label{l1closed}
\end{eqnarray}

\bigskip

\noindent The use of these formulas gives iso-$l_1$ lines on a $\Omega_M - \Omega_{\Lambda}$ plot
in $25 \sim 50 $\% agreement with the corresponding results found from computer code.
On the insensitivity of $l_1$ to other variables, see\cite{HW1,HW2}.
The derivation of these formulas was given in \cite{FNR}. Here we add some more details.

\bigskip

\noindent The formula for $l_1$ was derived from the relation $l_1 = \pi/\Delta\theta$ where
$\Delta\theta$ is the angle subtended by the horizon at the end of the recombination transition.
Let us consider the Legendre integral transform which has
as integrand a product of two factors, one
is the temperature autocorrelation function of the cosmic background
radiation and the other factor is a Legendre polynomial of degree $l$.
The issue is what is the lowest integer $l$ for which the two factors reinforce to create
the doppler peak? For small $l$ there is no reinforcement because the horizon
at recombination subtends a small angle about one degree
and the CBR fluctuations average to zero in the integral
of the Legendre transform.
At large $l$ the Legendre polynomial itself fluctuates with
almost equispaced nodes and antinodes.
The node-antinode spacing over which the Legendre polynomial varies
from zero to a local maximum in magnitude
is, in terms of angle, on average $\pi$ divided by $l$. When this angle coincides with the angle
subtended by the last-scattering horizon,
the fluctuations of the two integrand factors are, for the first time with increasing $l$,
synchronized and reinforce (constructive interference) and
the corresponding partial wave coefficient is larger than for slightly smaller
or slightly larger $l$.
This explains the occurrence of $\pi$ in the equation for the $l_1$ value of the first doppler peak
written as $l_1 = \pi/\Delta\theta$.

\bigskip

Another detail concerns whether to use the photon or
acoustic horizon, where the former is $\sqrt{3}$
larger than the latter?
If we examine the evolution
of the recombination transition given in \cite{Peebles}
the degree of ionization is 99\% at $5,000^0$K
(redshift $Z=1,850$) falling to 1\% at $3,000^0$K ($Z = 1,100$).
One can see qualitatively that
during the recombination transition the fluctuation
can grow.
The agreement
of the formula for $l_1$, using the photon horizon,
with experiment shows phenomenologcally
that the fluctuation does grow
during the recombination transition and that is why there is no full factor of
$\sqrt{3}$, as would arise using the acoustic horizon, in its
numerator.
When we look at the CMBFAST code below,
we shall find a factor
in $l_1$ of $\sim1.22$, intermediate between $1$ (optical)
and $\sqrt{3}$ (acoustic).

\bigskip
\bigskip

To introduce our quintessence model as a time-dependent cosmological term,
we start from the Einstein equation:

\begin{equation}
R_{\mu\nu} - \frac{1}{2} R g_{\mu\nu} = \Lambda(t) g_{\mu\nu} + 8 \pi G T_{\mu\nu} = 8 \pi G {\cal T}_{\mu\nu}
\label{einstein2}
\end{equation}

\noindent where $\Lambda(t)$ depends on time as will be specified later
and $T_{\nu}^{\mu} = {\rm diag} (\rho, -p, -p, -p)$.
Using the Robertson-Walker metric, the `00' component of Eq.(\ref{einstein2})
is

\begin{equation}
\left( \frac{\dot{R}}{R} \right)^2 + \frac{k}{R^2} = \frac{8 \pi G \rho}{3} +
 \frac{1}{3}\Lambda
\label{00}
\end{equation}

\noindent while the `ii' component is
\begin{equation}
2\frac{\ddot{R}}{R} + \frac{\dot{R}^2}{R^2} + \frac{k}{R^2} = -8 \pi G p + \Lambda
\label{ii}
\end{equation}
Energy-momentum conservation follows from Eqs.(\ref{00},\ref{ii}) because of the Bianchi identity
$D^{\mu} (R_{\mu\nu} - \frac{1}{2} g_{\mu\nu}) = D^{\mu} (\Lambda g_{\mu\nu} + 8\pi G T_{\mu\nu})
= D^{\mu} {\cal T}_{\mu\nu} = 0$.

\bigskip

Note that the separation of ${\cal T}_{\mu\nu}$ into two terms, one involving $\Lambda(t)$,
as in Eq(\ref{einstein}), is not meaningful except in a phenomenological sense because of energy conservation.

\bigskip

In the present cosmic era, denoted by the subscript `0', Eqs.(\ref{00},\ref{ii}) become respectively:

\begin{equation}
\frac{8\pi G}{3} \rho_0 = H_0^2 + \frac{k}{R_0^2} - \frac{1}{3} \Lambda_0
\label{00now}
\end{equation}
\begin{equation}
- 8 \pi G p_0 = - 2 q_0 H_0^2 + H_0^2 + \frac{k}{R_0^2} - \Lambda_0
\label{iinow}
\end{equation}
where we have used $q_0 = - \frac{\ddot{R}_0}{R_0 H_0^2}$ and $H_0 = \frac{\dot{R}_0}{R_0}$.

For the present era, $p_0 \ll \rho_0$ for cold matter and then Eq.(\ref{iinow}) becomes:

\begin{equation}
q_0 = \frac{1}{2} \Omega_{M} - \Omega_{\Lambda}
\label{decel}
\end{equation}
where $\Omega_{M} = \frac{8 \pi G \rho_0}{3 H_0^2}$ and $\Omega_{\Lambda} = \frac{\Lambda_0}{3 H_0^2}$.

\bigskip
\bigskip

Now we can introduce the form of $\Lambda(t)$ we shall assume by writing

\begin{equation}
\Lambda(t) = b R(t)^{-P}
\end{equation}
where $b$ {\bf is} a constant and the exponent $P$ we shall study for the
range $0 \leq P < 3$. This motivates the introduction of the new variables

\begin{equation}
\tilde{\Omega}_M = \Omega_M - \frac{P}{3 - P} \Omega_{\Lambda} , ~~~~\tilde{\Omega}_{\Lambda}
= \frac{3}{3 - P} \Omega_{\Lambda}
\label{tilde}
\end{equation}

\noindent It is unnecessary to redefine $\Omega_C$
because $\tilde{\Omega}_M + \tilde{\Omega}_{\Lambda}
= \Omega_M + \Omega_{\Lambda}$. The case $P=2$ was proposed, at least for late
cosmological epochs, in \cite{chen}.

\bigskip
\bigskip

The equations for the first Doppler peak incorporating the possibility of non-zero $P$
are found to be the following modifications of Eqs.(\ref{l1flat},\ref{l1open},\ref{l1closed}).
For $\Omega_C=0$

\bigskip

\begin{equation}
l_1 = \pi \left( \frac{R_t}{R_0} \right)
\left[\tilde{\Omega}_M  \left( \frac{R_0}{R_t} \right)^3 + \tilde{\Omega}_{\Lambda}
\left( \frac{R_0}{R_t} \right)^P
 \right]^{1/2}
\int_1^{\frac{R_0}{R_t}} \frac{dw}{\sqrt{\tilde{\Omega}_M w^3 + \tilde{\Omega}_{\Lambda} w^P}}
\label{l1Pflat}
\end{equation}

\bigskip

\noindent If $\Omega_{C} < 0$ the formula becomes

\bigskip

\begin{eqnarray}
l_1 & = & \frac {\pi}{\sqrt{-\Omega_C}} \left( \frac{R_t}{R_0} \right)
\left[\tilde{\Omega}_M  \left( \frac{R_0}{R_t} \right)^3 + \tilde{\Omega}_{\Lambda}
\left( \frac{R_0}{R_t} \right)^P
 + \Omega_C
\left( \frac{R_0}{R_t} \right)^2 \right]^{1/2} \times \nonumber \\
& & ~~~~ \times {\rm sin} \left( \sqrt{-\Omega_C} \int_1^{\frac{R_0}{R_t}} \frac{dw}{\sqrt{\tilde{\Omega}_M w^3
+ \tilde{\Omega}_{\Lambda} w^P + \Omega_C w^2}}
\right)
\label{l1Popen}
\end{eqnarray}

\bigskip

\noindent For the third possibility of a closed universe with $\Omega_C > 0$ the formula
is:

\bigskip

\begin{eqnarray}
l_1 & = & \frac {\pi}{\sqrt{\Omega_C}} \left( \frac{R_t}{R_0} \right)
\left[\tilde{\Omega}_M  \left( \frac{R_0}{R_t} \right)^3 + \tilde{\Omega}_{\Lambda}
\left( \frac{R_0}{R_t} \right)^P
 + \Omega_C
\left( \frac{R_0}{R_t} \right)^2 \right]^{1/2} \times \nonumber \\
& & ~~~ \times ~~~ {\rm sinh} \left( \sqrt{\Omega_C} \int_1^{\frac{R_0}{R_t}} \frac{dw}{\sqrt{\tilde{\Omega}_M w^
3
+ \tilde{\Omega}_{\Lambda} w^P + \Omega_C w^2}}
\right)
\label{l1Pclosed}
\end{eqnarray}

\bigskip
\bigskip

\noindent The dependence of $l_1$ on $P$ is illustrated for constant $\Omega_M = 0.3$ in
Fig. 1(a), and for the flat case $\Omega_C = 0$ in Fig. 1(b). (These figures
are in \cite{CDFNR}) For illustration we have varied
$0 \leq P < 3$ but as will become clear later in the paper (see Fig 3 
of \cite{CDFNR} discussed below) only the
much more restricted range $0 \leq P < 0.2$ is possible for a fully consistent
cosmology when one considers evolution since the
nucleosynthesis era.

\bigskip
\bigskip

\noindent We have introduced $P$ as a parameter which is real and with
$0 \leq P < 3$.
For $P \rightarrow 0$ we regain the standard cosmological model. But now
we must investigate other restrictions already necessary for $P$ before precision
cosmological measurements restrict its range even further.

\bigskip
\bigskip

Only for certain $P$ is it possible to extrapolate the cosmology
consistently for all $0 < w = (R_0/R) < \infty$. For example, in the flat case $\Omega_C = 0$
which our universe seems to approximate\cite{boomerang}, the formula for the expansion rate is

\begin{equation}
\frac{1}{H_0^2} \left( \frac{\dot{R}}{R} \right)^2 = \tilde{\Omega}_M w^3 + \tilde{\Omega}_{\Lambda} w^P
\label{flatexp}
\end{equation}

\noindent This is consistent as a cosmology only if the right-hand side has no zero for a real
positive $w = \hat{w}$. The root $\hat{w}$ is

\begin{equation}
\hat{w} = \left( \frac{ 3(1 - \Omega_M)}{P - 3 \Omega_M} \right)^{\frac{1}{3 - P}}
\label{wcrit}
\end{equation}

\bigskip

\noindent If $0 < \Omega_M < 1$, consistency requires that $P < 3 \Omega_M$.

\bigskip
\bigskip

\noindent In the more
general case of $\Omega_C \neq 0$ the allowed regions of the $\Omega_M - \Omega_{\Lambda}$
plot for $P = 0,1,2$ are displayed in Fig. 2 of \cite{CDFNR}.

\bigskip
\bigskip

\noindent We see from Eq.(\ref{wcrit}) that if we do violate $P < 3 \Omega_M$ for the flat case
then there is a $\hat{w} > 0$ where the cosmology undergoes a bounce, with $\dot{R} = 0$
and $\dot{R}$ changing sign. This necessarily arises
because of the imposition of $D^{\mu} {\cal T}_{\mu\nu} = 0$ for energy conservation. For this example
it occurs in the past for $\hat{w} > 1$.
The consistency of big bang cosmology back to the time of nucleosynthesis implies
that our universe has not bounced for any $1 < \hat{w} < 10^9$.
It is also possible to construct cosmologies where the bounce occurs in the future!
Rewriting Eq.(\ref{wcrit}) in terms of $\Omega_{\Lambda}$:

\begin{equation}
\hat{w} = \left( \frac{3 \Omega_{\Lambda}}{3 \Omega_{\Lambda} - (3 - P)} \right)^{\frac{1}{3 - P}}
\end{equation}

\bigskip

\noindent If $P < 3$, then any $\Omega_{\Lambda} < 0$ will lead to a solution with $0 < \hat{w} < 1$
corresponding to a bounce in the future. If $P > 3$ the condition for a future
bounce is $\Omega_{\Lambda} < - \left( \frac{P - 3}{3} \right)$. What this means is that
for the flat case $\Omega_C = 0$ with quintessence $P > 0$ it is
possible for the future cosmology to be qualitatively similar to a non-quintessence
closed universe where $\dot{R} = 0$ at a finite future time with a subsequent big crunch.

\bigskip
\bigskip

Another constraint on the cosmological model is provided by nucleosynthesis which requires that
the rate of expansion for very large $w$ does not differ too much from that
of the standard model.

\bigskip

The expansion rate for $P = 0$ coincides for large $w$ with that of the standard model so it
is sufficient to study the ratio:

\begin{eqnarray}
(\dot{R}/R)_P^2/(\dot{R}/R)_{P=0}^2
&
\stackrel{w \rightarrow \infty}{\rightarrow}
& (3 \Omega_{M} - P)/((3 - P) \Omega_{M})\\
& \stackrel{w \rightarrow \infty}{\rightarrow}& (4 \Omega_{R} - P)/((4 - P)\Omega_{R})
\end{eqnarray}

\noindent where the first limit is for matter-domination and the second is for radiation-domination
(the subscript R refers to radiation).

\bigskip

\noindent The overall change in the expansion rate at the BBN era is therefore

\begin{eqnarray}
(\dot{R}/R)_P^2/(\dot{R}/R)_{P=0}^2
& \stackrel{w \rightarrow \infty}{\rightarrow} &
(3 \Omega_{M} - P)/((3 - P) \Omega_{M}) \times \nonumber \\
& & \times ~~~ (4 \Omega^{trans}_{R} - P)/((4 - P)\Omega^{trans}_{R})
\end{eqnarray}
\noindent where the superscript ``trans" refers to the transition from radiation domination
to matter domination.
Putting in the values $\Omega_M = 0.3$ and $\Omega^{trans}_R = 0.5$ leads to $P < 0.2$
in order that the acceleration rate at BBN be within 15\% of its value in the standard model,
equivalent to the contribution to the expansion rate at BBN of one chiral neutrino flavor.

\bigskip
\bigskip

\noindent Thus the constraints of avoiding a bounce ($\dot{R} = 0$) in the past, and then requiring
consistency with BBN leads to $0 < P < 0.2$.

\bigskip

We may now ask how this restricted range of $P$ can effect the extraction
of cosmic parameters from observations. This demands an accuracy which has
fortunately begun to be attained with the Boomerang data\cite{boomerang}. If we choose
$l_1 = 197$ and vary $P$ as $P = 0, 0.05, 0.10, 0.15, 0.20$ we
find in the enlarged view of Fig 3 in \cite{CDFNR}
that the variation in the parameters
$\Omega_M$ and $\Omega_{\Lambda}$ can be as large as $\pm 3\%$. To guide
the eye we have added the line for deceleration parameter $q_0 = -0.5$ as suggested
by \cite{Perlmutter,Kirshner}. In the next
decade, inspired by the success of Boomerang (the first paper of true precision cosmology)
surely the sum $(\Omega_M + \Omega_{\Lambda})$ will be examined at much better than $\pm 1\%$
accuracy, and so variation of the exponent of $P$ will provide a useful parametrization
of the quintessence alternative to the standard cosmological model with constant $\Lambda$.

\bigskip

Clearly, from the point of view of inflationary cosmology, the precise
vanishing of $\Omega_C = 0$ is a crucial test and its confirmation
will be facilitated by comparison models such as the present one.

\bigskip

We have also studied the use of the public code CMBFAST\cite{CMBFAST}
and how its normalization compares to that in \cite{FNR}.
For example, with P = 0 and $\Omega_{\Lambda} = 0.3, h_{100} = 0.65$ we
find using CMBFAST that

\bigskip

$\Omega_{\Lambda} = 0.5, l_1 = 284$  ($l_1 = 233$ from \cite{FNR})

$\Omega_{\Lambda} = 0.6, l_1 = 254$  ($l_1 = 208$ from \cite{FNR})

$\Omega_{\Lambda} = 0.7, l_1 = 222$  ($l_1 = 182$ from \cite{FNR})

$\Omega_{\Lambda} = 0.8, l_1 = 191$  ($l_1 = 155$ from \cite{FNR})

\bigskip

The CMBFAST $l_1$ values are consistently $\sim 1.22$ times the $l_1$ values from \cite{FNR}.
As mentioned earlier, this normalization is intermediate between that for
the acoustic horizon ($\sqrt{3}$) and the
photon horizon ($1$).

\bigskip

Finally, we remark that the quintessence model considered here is in
the right direction to ameliorate the ``age
problem" of the universe. Taking the age as 14.5Gy for $\Omega_M = 0.3$,
$\Omega_C = 0$ and $h_{100} = 0.65$ the age increases monotonically with $P$.
It reaches slightly over         15 Gy at the highest-allowed value
$P= 0.2$. This behavior is illustrated in Fig 4 of \cite{CDFNR}
which assumes $\Omega_M = 0.3$ and flatness as $P$ is varied.

\bigskip
\bigskip

\begin{acknowledgments}
This work was supported in part by the US Department of Energy
under Grant No. DE-FG02-97ER-41036.
\end{acknowledgments}


\begin{chapthebibliography}{99}
\bibitem{int}
We note the similarity to the formula for the age of the universe:
\begin{equation}
t_0 = \frac{1}{H_0}\int_1^{\infty}\frac{dw}{w\sqrt{\Omega_{\Lambda}
+ \Omega_C w^2 + \Omega_0 w^3}}    \nonumber
\end{equation}
which is also an elliptic integral that cannot be done analytically.
\end{chapthebibliography}
\end{document}